\documentclass[aps,floats]{revtex4}
\usepackage{amsmath,amssymb}
\usepackage{graphicx,epsfig}
\usepackage[greek,english]{babel}

\begin{document}
\bibliographystyle {plain}

\def\oppropto{\mathop{\propto}} 
\def\opsimeq{\mathop{\simeq}}
\def\opoverderline{\mathop{\overline}}
\def\operarrow{\mathop{\longrightarrow}}
\def\opsim{\mathop{\sim}}

\def\fig#1#2{\includegraphics[height=#1]{#2}}
\def\figx#1#2{\includegraphics[width=#1]{#2}}


\title{ Periodically driven random quantum spin chains : \\
Real-Space Renormalization for Floquet localized phases  } 


\author{ C\'ecile Monthus }
 \affiliation{Institut de Physique Th\'{e}orique, 
Universit\'e Paris Saclay, CNRS, CEA,
91191 Gif-sur-Yvette, France}

\begin{abstract}
When random quantum spin chains are submitted to some periodic Floquet driving, the eigenstates of the time-evolution operator over one period can be localized in real space. For the case of periodic quenches between two Hamiltonians (or periodic kicks), where the time-evolution operator over one period reduces to the product of two simple transfer matrices, we propose a Block-self-dual renormalization procedure to construct the localized eigenstates of the Floquet dynamics. We also discuss the corresponding Strong Disorder Renormalization procedure, that generalizes the RSRG-X procedure to construct the localized eigenstates of time-independent Hamiltonians. 

\end{abstract}

\maketitle

\section{ Introduction  }

The issue of thermalization in isolated quantum many-body systems,
which has been much analyzed recently
 for time-independent Hamiltonians (see the reviews \cite{revue_eth,revue_huse} and references therein), has been also studied for time-periodic Hamiltonians 
\cite {prosen_kick98,rigol_floquetpure,moessner_floquetpure,prosen_kick14,roeck15,prosen_kick16,saito,mori,prosen_kick17,roeck17} : the usual decomposition of the unitary dynamics in terms of the eigenmodes of the time-independent-Hamiltonian
is then replaced by the decomposition into the eigenmodes of the time-evolution-operator over one period $T$ within the Hilbert space of size ${\cal N}$
\begin{eqnarray}
 U(T,0) \equiv {\cal T} e^{-i \int_0^T d t H(t) }  =  \sum_{n=1}^{\cal N}  e^{-i \theta_n}   \vert u_n \rangle \langle  u_n \vert
\label{Utimeorder}
\end{eqnarray}
The phases $\theta_n \in ]-\pi,+\pi]$ characterizing the eigenvalues $e^{-i \theta_n}  $ of this unitary operator are often rewritten as
\begin{eqnarray}
\theta_n= T \epsilon_n
\label{thetan}
\end{eqnarray}
where the Floquet quasi-energies $\epsilon_n$ are only defined modulo $\frac{2 \pi}{T}$.
The time-evolution-operator $U(T,0)$ can be then rewritten 
\begin{eqnarray}
U(T,0) = e^{-i T H_F }
\label{UHfloquet}
\end{eqnarray}
as if it were associated to the time-independent Floquet Hamiltonian
\begin{eqnarray}
H_F = \sum_{n=1}^{\cal N}  \epsilon_n  \vert u_n \rangle \langle  u_n \vert
\label{Hfloquet}
\end{eqnarray}
This construction in terms of the spectral decomposition of the evolution operator of Eq. \ref{Utimeorder} shows that the Floquet Hamiltonian 
associated to a finite period $T$ is usually very implicit in terms of the microscopic local degrees of freedom. 
However in the limit of short period $T \to 0$, 
the Floquet Hamiltonian is given by the high-frequency Magnus expansion
(see the review \cite{revue_magnus} and references therein)
\begin{eqnarray}
H_F && \opsimeq_{T \to 0}  H_{av} + H_2+H_3 +...
\nonumber \\
H_{av} && = \frac{1}{T} \int_0^T d t H(t) 
\nonumber \\
H_2 && = \frac{1}{2T} \int_0^T d t_1 \int_0^{t_1} dt_2 [ H(t_1),H(t_2)] 
\label{magnus}
\end{eqnarray}
where the leading term corresponds simply to the time-averaged Hamiltonian $H_{av}$, while all the other corrections involve various commutators.
Roughly speaking, this Magnus expansion
 is expected to converge as long as the phases $\theta_n^{av} = T E_n^{av}$ associated to the time-averaged Hamiltonian $H_{av}$ 
remain within the first Brillouin zone $-\pi \langle  \theta_n^{av} \langle \pi$ \cite{rigol_floquetpure}. However, in a many-body quantum system involving $N$ spins,
 the full bandwidth grows extensively $E_{max}^{av} - E^{av}_{min} \propto N$, and the typical energies grows as $E^{av}_{typ} \propto \sqrt {N} $,
so that the radius of convergence of the Magnus expansion $T \leq T_c(N) $ shrinks to zero in the thermodynamic limit $T_c(N \to +\infty) \to 0$ \cite{rigol_floquetpure}.
For generic non-integrable extensive systems driven with a finite period $T$, the Magnus expansion thus breaks down and the Floquet dynamics leads
 to the Random Matrix Circular Ensemble statistics 
\cite{rigol_floquetpure} for the Floquet phases $\theta_n$
that are uniformly distributed on $]-\pi,+\pi]$ :  the existing level repulsion can be interpreted as resulting from the strong mixing between the energy levels of the undriven system that correspond to the same quasi-energy defined modulo $\frac{2 \pi}{T}$ \cite{moessner_floquetpure}.
However this general conclusion about thermalization can be avoided in the presence of disorder if the Floquet eigenstates are localized in real space
 \cite{ponte_local,ponte_global,abanin_arxiv,lazarides,rehn,dmrgfloquet,andersonfloquet,khemani,timecrystal,yao,shortrev_phasesfloquet}, in analogy
 with the phenomena of Anderson Localization or Many-Body-Localization
for time-independent random Hamiltonians. In addition, when the Floquet eigenstates are localized in real space, they may display
different types of order, like Spin-Glass or Paramagnetic as for time-independent problems, but also new phases specific to the Floquet 
periodic driving that involve for instance the period-doubling phenomenon for some degrees of freedom \cite{khemani,timecrystal,yao,shortrev_phasesfloquet}.

The aim of the present paper is to introduce real-space renormalization procedures for the Floquet localized eigenstates of random quantum spin chains.
The paper is organized as follows.
In section \ref{sec_model}, we describe the model of a random quantum spin chain submitted to periodic quenches between two Hamiltonians
or to periodic kicks. In section \ref{sec_rgmatrix}, we derive the block real space renormalization rules for the parameters of the evolution operator
over one period. In section \ref{sef_rgrules}, we discuss the properties of these Block-RG rules, and describe the alternative Strong Disorder RG procedure.
Our conclusions are summarized in section \ref{sec_conclusion}.

\section{ Random quantum spin chain submitted to periodic quenches or kicks  }

\label{sec_model}

\subsection { Periodic sudden quenches between two Hamiltonians }

In this paper, we consider the periodic Hamiltonian $H(t+T)=H(t) $
of period $T=T_0+T_1$ with the following dynamics over one period
\begin{eqnarray}
H(0 \leq t \leq T_0)  && =H_0 \equiv  - \sum_{n=1}^{N-1} J_n \sigma^z_n \sigma^z_{n+1}
\nonumber \\
H(T_0 \leq t \leq T=T_0+T_1)  && =H_1 \equiv  - \sum_{n=1}^N  h_n \sigma_n^x
\label{h0h1}
\end{eqnarray}
The evolution operator during the period reads
\begin{eqnarray}
U(t,0) &&=  e^{-i t H_0 } = e^{\displaystyle i t \sum_{n=1}^{N-1} J_n \sigma^z_n \sigma^z_{n+1}} \ \ \ \ \ \ \ \ \ \ \ \  \ {\rm for } \ \ 0 \leq t \leq T_0
\nonumber \\
U(t, T_0) && = e^{-i (t-T_0) H_1 }  =e^{\displaystyle i (t-T_0) \sum_{n=1}^N  h_n \sigma_n^x } \ \ \ \ \ \ {\rm for } \ \ T_0 \leq t \leq T= T_0+T_1
\label{Uperiod}
\end{eqnarray}
In particular, the evolution operator over one period 
\begin{eqnarray}
U^{cycle} \equiv U(T,0) = U(T=T_0+T_1,T_0) U(T_0,0) = e^{-i T_1 H_1 } e^{-i T_0 H_0 } = e^{\displaystyle i T_1 \sum_{n=1}^N  h_n \sigma_n^x } \ \ 
 e^{\displaystyle i T_0 \sum_{n=1}^{N-1} J_n \sigma^z_n \sigma^z_{n+1}} 
\label{Ucycle}
\end{eqnarray}
is explicit as the product of the two evolution operators associated to the two Hamiltonians $(H_0,H_1)$.
This type of periodic sudden quenches between two Hamiltonians 
is thus much simpler technically than the general case of a continuously-varying
Hamiltonian where the evolution operator requires the time-ordering of Eq. \ref{Utimeorder}. This explains why this protocol of periodic quenches
 has been the most considered framework recently in the field of random quantum spin chains \cite{ponte_local,ponte_global,dmrgfloquet,khemani,timecrystal,yao,shortrev_phasesfloquet}.

\subsection { Periodically-kicked Quantum spin chain}

The periodically kicked spin chain of Hamiltonian of period $T_0$
\begin{eqnarray}
H(\tau)  && =H_0 + H_{kick} \sum_{m=-\infty}^{+\infty}  \delta(\tau-m T_0)
\label{hkicked}
\end{eqnarray}
yields the cyclic evolution operator 
\begin{eqnarray}
U^{cycle}  && =e^{-i H_{kick} } e^{-i T_0 H_0 } 
\label{ukicked}
\end{eqnarray}
that coincides with  Eq. \ref{Ucycle} with the correspondence $H_{kick} \to T_1 H_1$.
This equivalent formulation makes the link with the literature on kicked quantum models which have a long history in
the field of quantum chaos (see \cite{prosen_kick98,prosen_kick14,prosen_kick16,prosen_kick17} and references therein).

\subsection { New possible phases in Floquet systems }

The evolution operator of Eq. \ref{Ucycle} is directly related via analytic continuation to
the Transfer Matrix of the two-dimensional classical Ising model with columnar disorder known as the McCoy-Wu model.
For the case without disorder, this link 
has been recently discussed in the context of integrable Floquet systems \cite{integrablefloquet}.
However, the Floquet dynamics involving phases introduces some differences with respect to statistical models involving real Boltzmann weights.

Indeed, the identities for the elementary transfer matrices
\begin{eqnarray}
  e^{i T_1 h_n \sigma_n^x } && = \cos( T_1 h_n) +i  \sin( T_1 h_n)  \sigma_n^x 
\nonumber \\
 e^{i T_0 J_n \sigma_n^z \sigma_{n+1}^z}&& = \cos( T_0 J_n) +i  \sin( T_0 J_n)  \sigma_n^z  \sigma_{n+1}^z
\label{Uelem}
\end{eqnarray}
show that the couplings $h_n$ and $J_n$ actually only appear via the cosinus and sinus of the phases $T_1 h_n $ and $T_0 J_n $, with a periodicity of $2 \pi$. In addition,  the consideration of the special values $(0, \pm \frac{\pi}{2},\pi)$
yields the possibility of new phases specific to the Floquet 
periodic driving  \cite{khemani,timecrystal,yao,shortrev_phasesfloquet}.
For instance, if all the transverse fields $h_n$ take the same non-random value,
besides the usual Spin-Glass case corresponding to $h_n=0$ 
\begin{eqnarray}
U_{h_n=0}^{cycle} = e^{\displaystyle i T_0 \sum_{n=1}^{N-1} J_n \sigma^z_n \sigma^z_{n+1}} 
\label{Uhnzero}
\end{eqnarray}
there exists a new $\pi$-Spin-Glass case corresponding to $h_n=\frac{\pi}{2 T_1}$
\begin{eqnarray}
U_{h_n=\frac{\pi}{2 T_1}}^{cycle} = i^N \left( \prod_n \sigma_n^x  \right) \ \ 
 e^{\displaystyle i T_0 \sum_{n=1}^{N-1} J_n \sigma^z_n \sigma^z_{n+1}} 
\label{Uhnpisur2}
\end{eqnarray}
where the states $\vert S_1,..,S_N\rangle$ in the $\sigma^z$ basis are eigenstates over two periods, but not over one period
as a consequence of the global flip of all the spins $\left( \prod_n \sigma_n^x  \right) $.

Similarly, if all the couplings $J_n$ take the same non-random value,
besides the usual Paramagnetic case corresponding to $J_n=0$ 
\begin{eqnarray}
U_{J_n=0}^{cycle}=  e^{\displaystyle i T_1 \sum_{n=1}^N  h_n \sigma_n^x } 
\label{Ujnzero}
\end{eqnarray}
there exists a new $0\pi$-Paramagnetic case corresponding to $J_n=\frac{\pi}{2 T_0}$
\begin{eqnarray}
U_{J_n=\frac{\pi}{2 T_0}}^{cycle} && =i^N e^{\displaystyle i T_1 \sum_{n=1}^N  h_n \sigma_n^x } \ \ 
 \prod_{n=1}^N (  \sigma^z_n \sigma^z_{n+1} )
=i^N e^{\displaystyle i T_1 \sum_{n=1}^N  h_n \sigma_n^x } \ \ 
   \sigma^z_1 \sigma^z_{N} 
\label{Ujnpisur2}
\end{eqnarray}
where the states in the $\sigma^x$ basis are eigenstates over two periods, but not over one period
as a consequence of the flip of the two end spins $\sigma^z_1 \sigma^z_{N} $.

\subsection{ Discussion }

The two examples above of the $\pi$-Spin-Glass and of the $0\pi$-Paramagnet \cite{khemani,timecrystal,yao,shortrev_phasesfloquet}
show that the description of the full general case where $T_1 h_n $ and $T_0 J_n $ can be random anywhere on $]-\pi,+\pi]$
requires the discussion of many separate cases to identify the degrees of freedom submitted to the period doubling phenomenon.
In the following, to simplify the discussion, we will thus focus on the case without any period doubling
where the random elementary phases remain in the sector
\begin{eqnarray}
 - \frac{\pi}{4} \leq  T_1 h_n \leq \frac{\pi}{4}
\nonumber \\
 - \frac{\pi}{4} \leq T_0 J_n \leq \frac{\pi}{4}
\label{sectorpisur4}
\end{eqnarray}

In the limit of short period $T \to 0$ where all these phases become infinitesimal,
the averaged Hamiltonian of the Magnus expansion of Eq. \ref{magnus}
for the protocol of Eq. \ref{h0h1}
\begin{eqnarray}
H_{av} && = \frac{1}{T} \int_0^T d t H(t) = - \sum_{n=1}^{N-1} J_n^{tav} \sigma^z_n \sigma^z_{n+1}  - \sum_{n=1}^N  h_n^{tav} \sigma_n^x
\label{havquantumising}
\end{eqnarray}
corresponds to the quantum Ising chain with the random couplings 
\begin{eqnarray}
J_n^{tav}  && \equiv \frac{ T_0 J_n }{T} 
\label{jnav}
\end{eqnarray}
and the random transverse fields
\begin{eqnarray}
h_n^{tav} && \equiv \frac{T_1 h_n}{T} 
\label{hnav}
\end{eqnarray}
The construction of its ground state by Daniel Fisher \cite{fisher} via the Strong Disorder RG approach (reviewed in \cite{strong_review}) has proven to be be extremely powerful to compute the critical properties of the 
quantum paramagnetic/Spin-Glass phase transition at zero temperature.
This approach has been later extended into the RSRG-X procedure
in order to construct the whole set of excited eigenstates of various random quantum spin chains in their localized phases
 \cite{rsrgx,rsrgx_moore,vasseur_rsrgx,yang_rsrgx,rsrgx_bifurcation}.
Another possibility to construct the whole set of eigenstates is the Block Self-Dual Renormalization procedure \cite{c_emergent}
that generalizes the Fernandez-Pacheco procedure to construct the ground state of the pure chain \cite{pacheco,igloiSD}
(see also the extensions to higher dimensions \cite{epele,nishiPur,kubica} or other models \cite{horn,hu,solyom,solyom2})
and of the random chain \cite{nishiRandom,us_pacheco,us_renyi,us_watermelon}. 
In the next section, our goal is to extend the idea of this Block Self-Dual Renormalization procedure
to the evolution operator of the Floquet dynamics when the elementary phases 
are finite in the sector of Eq \ref{sectorpisur4}
instead of infinitesimal (Eq. \ref{havquantumising}).
The corresponding Strong Disorder RG procedure will be discussed in section \ref{sec_strong}.

But before we enter the RG technicalities, it seems useful to stress the following points :

(i) the Floquet model of Eq. \ref{h0h1} that we consider can be mapped onto free-fermions via the Jordan-Wigner transformation,
so that the phenomenon of thermalization discussed in the Introduction can never occur here. In addition, 
 the presence of disorder in this one-dimensional model leads to the Anderson-real-space-localization of all fermionic modes
(instead of real-space-delocalized Fourier modes for the translation-invariant model without disorder).

(ii) in terms of the spin degrees of freedom, 
the corresponding localized Floquet eigenstates may display different types of order, namely Paramagnetic and Spin-Glass
if one focus on the sector of Eq. \ref{sectorpisur4} to avoid the discussion of the possible period-doubling phenomena recalled above.
The goal of the RG procedure described below is to characterize the phase transition between these Paramagnetic
 and Spin-Glass localized phases
of the Floquet model, and to discuss the link with the Fisher Strong Disorder RG describing the Paramagnetic/Spin-Glass
transition of the ground-state of the quantum Ising chain of Eq. \ref{havquantumising}.

\section{ Block Self-dual Renormalization procedure }

\label{sec_rgmatrix}

In this section, we describe how the idea of the Block Self-Dual Renormalization procedure \cite{pacheco,igloiSD,nishiRandom,us_pacheco,us_renyi,us_watermelon,c_emergent}
concerning the random quantum Ising chain of Eq. \ref{havquantumising} 
can be adapted to the evolution operator of the Floquet dynamics of Eq. \ref{Ucycle}.

\subsection{ Transfer matrices associated to even-odd couplings and fields}

To keep the duality between couplings and transverse fields during the renormalization, it is convenient to separate even and odd couplings and fields
within the evolution operator (see \cite{cardy} for the case without disorder
in the language of the transfer matrix for the two-dimensional classical Ising model)
\begin{eqnarray}
U^{cycle}  = e^{ \displaystyle iT_1 \sum_n \  h_{2n} \sigma_{2n}^x }
\ \ e^{ \displaystyle iT_1 \sum_n  h_{2n-1}  \sigma_{2n-1}^x }
\ \ e^{ \displaystyle iT_0 \sum_n J_{2n-1} \sigma^z_{2n-1} \sigma^z_{2n} }
\ \ e^{ \displaystyle iT_0 \sum_n J_{2n} \sigma^z_{2n} \sigma^z_{2n+1} }
\label{Uevenodd}
\end{eqnarray}
in order to introduce
\begin{eqnarray}
{\cal M}  = 
\ \ e^{ \displaystyle iT_1 \sum_n  h_{2n-1}  \sigma_{2n-1}^x }
\ \ e^{ \displaystyle iT_0 \sum_n J_{2n-1} \sigma^z_{2n-1} \sigma^z_{2n} }
\label{defM}
\end{eqnarray}
and
\begin{eqnarray}
{\cal N} = 
 e^{ \displaystyle iT_0 \sum_n J_{2n} \sigma^z_{2n} \sigma^z_{2n+1} }
\ \ e^{ \displaystyle iT_1 \sum_n \  h_{2n} \sigma_{2n}^x }
\label{defN}
\end{eqnarray}
Then Eq. \ref{Uevenodd} can be rewritten as 
\begin{eqnarray}
U^{cycle}  = 
e^{ - \displaystyle iT_0 \sum_n J_{2n} \sigma^z_{2n} \sigma^z_{2n+1} } {\cal N} {\cal M} e^{ \displaystyle iT_0 \sum_n J_{2n} \sigma^z_{2n} \sigma^z_{2n+1} }
\label{Uevenoddmn}
\end{eqnarray}
so that the evolution over $p$ cycles
\begin{eqnarray}
U(p T,0) = [ U(T,0]^p =  e^{ - \displaystyle iT_0 \sum_n J_{2n} \sigma^z_{2n} \sigma^z_{2n+1} }
\ \  ( {\cal N} {\cal M})^p \ \  e^{ \displaystyle iT_0 \sum_n J_{2n} \sigma^z_{2n} \sigma^z_{2n+1} }
\label{Uoneperiodmn}
\end{eqnarray}
involves the alternate product of the matrices $ {\cal N}$ and ${\cal M} $ up to boundary terms.

\subsection{ Spectral analysis of the matrix ${\cal M}$ }

The matrix ${\cal M}$ of Eq. \ref{defM} commutes with all $\sigma^z_{2n}$.
As a consequence,
the matrix elements in the $\sigma^z$ basis can be factorized into independent one-spin problems for the odd spins
$\sigma_{2n-1}$
\begin{eqnarray}
\langle  S_1',...,S_{N}' \vert {\cal M}  \vert S_1,...,S_N \rangle= \prod_{n=1}^{\frac{N}{2}} 
\delta_{S_{2n}',S_{2n} } 
\langle S_{2n-1}' \vert e^{iT_1   h_{2n-1}  \sigma_{2n-1}^x }  e^{iT_0  J_{2n-1} \sigma^z_{2n-1} S_{2n} }  \vert S_{2n-1} \rangle
\label{Melements}
\end{eqnarray}

So for each value $S_{2n}=\pm 1$, one has to diagonalize 
the two-by-two unitary matrix concerning the single quantum spin $\sigma_{2n-1} $
\begin{eqnarray}
{\cal M}_{2n-1;S_{2n}} && =
e^{iT_1 h_{2n-1} \sigma_{2n-1}^x} e^{ iT_0 J_{2n-1} S_{2n}  \sigma_{2n-1}^z   } 
= \left[  \cos (T_1 h_{2n-1} ) + i \sin(T_1 h_{2n-1}) \sigma_{2n-1}^x\right] e^{ iT_0 J_{2n-1} S_{2n}  \sigma_{2n-1}^z   } 
\nonumber \\
&& = \begin{pmatrix}
\cos(T_1 h_{2n-1} ) e^{ iT_0J_{2n-1} S_{2n}} &  i\sin(T_1 h_{2n-1}) e^{- iT_0J_{2n-1} S_{2n}} \\
i \sin(T_1 h_{2n-1}) e^{ iT_0J_{2n-1} S_{2n}} & \cos(T_1 h_{2n-1} ) e^{- iT_0J_{2n-1} S_{2n}}
\end{pmatrix}
\label{calmteff}
\end{eqnarray}

It is convenient to introduce the notations
\begin{eqnarray}
r_{2n-1} \equiv \frac{1}{ \sqrt{1+ \frac{\tan^2 (T_1 h_{2n-1}) }{ \sin^2 ( T_0 J_{2n-1}) } }}
\nonumber \\
\eta_{2n-1} \equiv {\rm sgn } \left(  \frac{\tan (T_1 h_{2n-1}) }{ \sin ( T_0 J_{2n-1}) } \right)
\label{reta}
\end{eqnarray}

The two eigenvalues are independent of the value of $S_{2n}=\pm 1$ and are complex-conjugate on the unit circle
\begin{eqnarray}
\lambda_{2n-1}^{\pm} =  \cos (T_1 h_{2n-1}) \left[ \cos ( T_0 J_{2n-1}) \pm i \frac{  \sin ( T_0 J_{2n-1})}{r_{2n-1} }  \right]
= e^{ \pm i \alpha_{2n-1} }
\label{calmeigen}
\end{eqnarray}
The unitary two-by-two matrix of Eq. \ref{calmteff} for the spin $\sigma_{2n-1} $ can be rewritten as the spectral decomposition
\begin{eqnarray}
{\cal M}_{2n-1;S_{2n}} &&  =\lambda^{+}_{2n-1} \vert \lambda^{+}_{2n-1}(S_{2n})\rangle\langle  \lambda^{+}_{2n-1}(S_{2n}) \vert
+ \lambda^{-}_{2n-1} \vert \lambda^{-}_{2n-1}(S_{2n})\rangle \langle  \lambda^{-}_{2n-1}(S_{2n}) \vert 
\label{calmteffspectral}
\end{eqnarray}
where the eigenvectors read
\begin{eqnarray}
 \vert \lambda^{+}_{2n-1}(S_{2n})\rangle&&
= e^{- \frac{iT_0J_{2n-1} }{2}S_{2n} } \sqrt{\frac{1+r_{2n-1} S_{2n}}{2} } \vert S_{2n-1}=+ \rangle+
\eta_{2n-1}  e^{ \frac{iT_0J_{2n-1} }{2}S_{2n} }\sqrt{\frac{1-r_{2n-1} S_{2n}}{2} } \vert S_{2n-1}=- \rangle
\label{lrp}
\nonumber
\end{eqnarray}
and
\begin{eqnarray}
 \vert \lambda^{-}_{2n-1}(S_{2n})\rangle&&=
- \eta_{2n-1} e^{- \frac{iT_0J_{2n-1} }{2}S_{2n} } \sqrt{\frac{1-r_{2n-1} S_{2n}}{2} }  \vert S_{2n-1}=+ \rangle
+  e^{ \frac{iT_0J_{2n-1} }{2}S_{2n} } \sqrt{\frac{1+r_{2n-1} S_{2n}}{2} }    \vert S_{2n-1}=- \rangle
\label{lrm}
\nonumber
\end{eqnarray}

For the pair $(\sigma_{2n-1},\sigma_{2n})$, there are thus two degenerate states associated to $\lambda^{+}_{2n-1} =  e^{  i \alpha_{2n-1} }   $
and two degenerate states associated to $\lambda^{-}_{2n-1}  = e^{ - i  \alpha_{2n-1} }  $ with the spectral decomposition
\begin{eqnarray}
{\cal M}_{2n-1}
&& =  e^{  i  \alpha_{2n-1} }  \sum_{S_{2n}=\pm 1} 
\vert \lambda^{+}_{2n-1}(S_{2n})\rangle\otimes \vert S_{2n} \rangle \langle  \lambda^{+}_{2n-1}(S_{2n}) \vert
\otimes \langle  S_{2n} \vert
\nonumber \\
&& +   e^{ - i  \alpha_{2n-1} } \sum_{S_{2n}=\pm 1}  
\vert \lambda^{-}_{2n-1}(S_{2n})\rangle\otimes \vert S_{2n} \rangle \langle  \lambda^{-}_{2n-1}(S_{2n}) \vert 
\otimes \langle  S_{2n} \vert
\label{projm}
\end{eqnarray}

The full matrix ${\cal M }  $ obtained by the product over all pairs
\begin{eqnarray}
{\cal M } && = \prod_{n=1}^{\frac{N}{2}} {\cal M}_{2n-1}
 =\prod_{n=1}^{\frac{N}{2}}  \left[ \sum_{\tau_{2n-1}=\pm 1}  e^{  i \tau_{2n-1} \alpha_{2n-1} } 
 \sum_{S_{2n}=\pm 1} 
\vert \lambda^{\tau_{2n-1}}_{2n-1}(S_{2n})\rangle\otimes \vert S_{2n} \rangle 
\langle  \lambda^{\tau_{2n-1}}_{2n-1}(S_{2n}) \vert \otimes \langle  S_{2n} \vert
 \right]
\label{mtspec}
\end{eqnarray}
is a transfer matrix of size $2^N \times 2^N$. Its $2^{\frac{N}{2}}$ eigenvalues
are labelled by the sequence of $\frac{N}{2}$ indices  $\tau_{2n-1}=\pm 1$
\begin{eqnarray}
M^{(\tau_1,.. ,\tau_{2n-1},.. )} =  e^{i \displaystyle \sum_{n=1}^{\frac{N}{2}} \tau_{2n-1} \alpha_{2n-1}    }
\label{meigenvalue}
\end{eqnarray} 
Each of these eigenvalues is degenerate $2^{\frac{N}{2}}$  times.
One basis of the corresponding degenerate subspace is given by the sequence of the $\frac{N}{2}$ values $S_{2n}$ of the even spins,
with the eigenvectors given by the tensor-products
\begin{eqnarray}
\vert M^{(\tau_1,.. ,\tau_{2n-1},.. )  }_{S_2,..,S_{2n}...}  \rangle =  \otimes_{n=1}^{\frac{N}{2}}  \vert \lambda^{\tau_{2n-1}}_{2n-1}(S_{2n})\rangle\otimes \vert S_{2n} \rangle
\label{meigenvectors}
\end{eqnarray} 

In the next section, the matrix ${\cal N}$ will be taken into account to lift this degeneracy and obtain the effective renormalized couplings and fields
for the even spins once the odd spins have been eliminated.

\subsection{ Matrix ${\cal N}$ between two matrices ${\cal M}$ }

Let us now focus on the matrix elements of the matrix of Eq. \ref{defN}
\begin{eqnarray}
{\cal N }&& =  e^{ \displaystyle iT_0 \sum_n J_{2n} \sigma^z_{2n} \sigma^z_{2n+1} }
\ \ e^{ \displaystyle iT_1 \sum_n \  h_{2n} \sigma_{2n}^x }
\label{nt}
\end{eqnarray}
within the degenerate subspace associated to each eigenvalue (Eq. \ref{meigenvalue}) and the corresponding basis of eigenvectors (Eq. \ref{meigenvectors}).
These matrix elements factorize into
\begin{eqnarray}
&& \langle  M^{(\tau_1,.. ,\tau_{2n-1},.. )  }_{S_2,..,S_{2n}...}  \vert  {\cal N }  \vert M^{(\tau_1,.. ,\tau_{2n-1},.. )  }_{S_2',..,S_{2n}'...}  \rangle   
 \nonumber \\ && 
= \prod_{n=1}^{\frac{N}{2}} 
\left[ \langle  \lambda^{\tau_{2n-1}}_{2n-1}(S_{2n}) \vert   e^{  iT_0  J_{2n-2} S_{2n-2} \sigma^z_{2n-1} }   \vert \lambda^{\tau_{2n-1}}_{2n-1}(S_{2n}')\rangle
\langle S_{2n} \vert e^{  iT_1  \  h_{2n} \sigma_{2n}^x }  \vert S_{2n}' \rangle  \right]
\label{nmatrix}
\end{eqnarray} 
so we need
\begin{eqnarray}
 \langle  S_{2n}\vert  e^{iT_1 h_{2n} \sigma_{2n}^x}  \vert S_{2n}'\rangle
&& = \langle  S_{2n} \vert  \left[  \cos( T_1 h_{2n} )+i \sin(T_1 h_{2n} ) \sigma_{2n}^x  \right] \vert S_{2n}'\rangle
\nonumber \\
&& = \cos( T_1 h_{2n} ) \delta_{S_{2n},S_{2n}' } +i \sin(T_1 h_{2n} )\delta_{S_{2n},-S_{2n}' } 
\label{nmatrix1}
\end{eqnarray}
and
\begin{eqnarray} 
&& \langle \lambda^{\tau_{2n-1}}_{2n-1} (S_{2n}) \vert  e^{ iT_0 J_{2n-2} S_{2n-2} \sigma_{2n-1}^z  } 
  \vert \lambda^{\tau_{2n-1}}_{2n-1}(S_{2n}')\rangle
 \nonumber \\
&& = \langle \lambda^{\tau_{2n-1}}_{2n-1} (S_{2n}) \vert
  \left[      \cos( T_0 J_{2n-2} ) + i \sin ( T_0 J_{2n-2})  S_{2n-2} \sigma_{2n-1}^z     \right]
 \vert \lambda^{\tau_{2n-1}}_{2n-1}(S_{2n}')\rangle
\label{nmatrix2}
\end{eqnarray}
Using the explicit expressions of the eigenvectors given in the previous subsection,
one obtains respectively for equal spins $S_{2n}=S_{2n}'$
\begin{eqnarray} 
  \langle \lambda^{\tau_{2n-1}}_{2n-1} (S_{2n}) \vert  e^{ iT_0 J_{2n-2} S_{2n-2} \sigma_{2n-1}^z  }   \vert \lambda^{\tau_{2n-1}}_{2n-1}(S_{2n}'=S_{2n})\rangle
=  \cos( T_0 J_{2n-2} )  +  i \sin ( T_0 J_{2n-2}) r_{2n-1} \tau_{2n-1}  S_{2n-2}  S_{2n}
\label{ntsame}
\end{eqnarray}
and for opposite spins
\begin{eqnarray} 
 && \langle \lambda^{\tau_{2n-1}}_{2n-1} (S_{2n}) \vert  e^{ iT_0 J_{2n-2} S_{2n-2} \sigma_{2n-1}^z  }   \vert \lambda^{\tau_{2n-1}}_{2n-1}(S_{2n}'=-S_{2n})\rangle
 \nonumber \\ &&
 =\sqrt{1-r_{2n-1}^2 } \left[  \cos( T_0 J_{2n-2} )  \cos(T_0J_{2n-1}  ) - \sin ( T_0 J_{2n-2})     \sin(T_0J_{2n-1}  )  S_{2n-2}  S_{2n} \right]
\label{ntopposite}
\end{eqnarray}

Putting everything together, the matrix elements within a degenerate subspace read
\begin{eqnarray}
&& \langle  M^{(\tau_1,.. ,\tau_{2n-1},.. )  }_{S_2,..,S_{2n}...}  \vert  {\cal N }  \vert M^{(\tau_1,.. ,\tau_{2n-1},.. )  }_{S_2',..,S_{2n}'...}  \rangle   
 \nonumber \\ && 
= \prod_{n=1}^{\frac{N}{2}} \cos( T_0 J_{2n-2} ) \cos( T_1 h_{2n} ) 
[ 
  \delta_{S_{2n},S_{2n}' } \left( 1 +  i \tan ( T_0 J_{2n-2}) r_{2n-1} \tau_{2n-1}  S_{2n-2}  S_{2n} \right)
 \nonumber \\ && 
+\delta_{S_{2n},-S_{2n}' }  i \sqrt{1-r_{2n-1}^2 }\tan(T_1 h_{2n} ) \cos(T_0J_{2n-1}  ) \left(  1- \tan ( T_0 J_{2n-2})     \tan(T_0J_{2n-1}  )  S_{2n-2}  S_{2n}   \right) ]
\label{ntfinal}
\end{eqnarray} 

To obtain a closed renormalization procedure that can be iterated, we wish to 
interpret these matrix elements as associated to some renormalized matrix 
displaying the same form as Eq \ref{nt} but involving only even spins $\sigma_{2n}$
and the appropriate renormalized couplings $J^R_{2n} $ and renormalized transverse fields $h^R_{2n} $
\begin{eqnarray}
{\cal N }^R&& =  e^{ \displaystyle iT_0 \sum_n J^R_{2n-2,2n} \sigma^z_{2n-2} \sigma^z_{2n} }
\ \ e^{ \displaystyle iT_1 \sum_n \  h^R_{2n} \sigma_{2n}^x }
\nonumber \\
&& = \left[ \prod_n \cos(T_0  J^R_{2n-2,2n} ) \left( 1+i \tan (T_0  J^R_{2n-2,2n}) \sigma^z_{2n-2} \sigma^z_{2n}   \right) 
\right]
\left[ \prod_n \cos(T_1  h^R_{2n} ) \left( 1+i \tan (T_1  h^R_{2n}) \sigma^x_{2n}   \right) 
\right]
\label{nr}
\end{eqnarray}
Let us now compare  Eq. \ref{ntfinal}
with the matrix elements
\begin{eqnarray}
&& \langle  S_2,..,S_{2n}...  \vert  {\cal N }^R  \vert S_2',..,S_{2n}'... \rangle   
\nonumber \\
&& = \left[ \prod_n \cos(T_0  J^R_{2n-2,2n} ) \left( 1+i \tan (T_0  J^R_{2n-2,2n}) S_{2n-2} S_{2n}   \right) 
\right]
\left[ \prod_n \cos(T_1  h^R_{2n} ) \left( \delta_{S_{2n},S_{2n}' }+i \tan (T_1  h^R_{2n})  \delta_{S_{2n},-S_{2n}' }  \right) 
\right]
\nonumber \\
&& = \prod_n \cos(T_0  J^R_{2n-2,2n} ) \cos(T_1  h^R_{2n} ) 
[ \delta_{S_{2n},S_{2n}' }\left( 1+i \tan (T_0  J^R_{2n-2,2n}) S_{2n-2} S_{2n}   \right) 
\nonumber \\
&& 
+  \delta_{S_{2n},-S_{2n}' } i \tan (T_1  h^R_{2n})  \left( 1+i \tan (T_0  J^R_{2n-2,2n}) S_{2n-2} S_{2n}   \right)  ]
\label{nrme}
\end{eqnarray}
once we have taken $\delta_{S_{2n},S_{2n}' } $ with coefficient unity as the reference term :
the best choice consists in matching the term in $ \delta_{S_{2n},S_{2n}' } S_{2n-2} S_{2n} $ yielding
 the renormalized couplings $  J^R_{2n-2,2n}$ between even spins $(\sigma_{2n-2}^z,\sigma_{2n}^z)$ 
\begin{eqnarray}
\tan(T_0 J^R_{2n-2,2n}) && = \tau_{2n-1}  \tan(T_0 J_{2n-2} ) r_{2n-1} 
\nonumber \\
&& = \tau_{2n-1}   \frac{\tan(T_0 J_{2n-2} )   \vert  \tan ( T_0 J_{2n-1})  \vert }
{ \sqrt{\tan^2 ( T_0 J_{2n-1})+\tan^2 (T_1 h_{2n-1}) +  \tan^2 (T_1 h_{2n-1}) \tan^2 ( T_0 J_{2n-1})  }}
\label{jr}
\end{eqnarray}
and in matching the term in $ \delta_{S_{2n},-S_{2n}' }  $
yielding the renormalized transverse fields $h^R_{2n} $ on the even spins
\begin{eqnarray}
\tan(T_1 h^R_{2n}) && =   \tan(T_1 h_{2n} )    \cos ( T_0 J_{2n-1} )
 \sqrt{ 1-   r_{2n-1}^2}  
\nonumber \\
&& =  
\frac{ \tan(T_1 h_{2n} )   \vert  \tan (T_1 h_{2n-1}) \vert }
{ \sqrt{\tan^2 ( T_0 J_{2n-1})+\tan^2 (T_1 h_{2n-1}) +  \tan^2 (T_1 h_{2n-1}) \tan^2 ( T_0 J_{2n-1})  }}
\label{hr}
\end{eqnarray}
while the terms in $\delta_{S_{2n},-S_{2n}' }  S_{2n-2} S_{2n} $ that would correspond to higher couplings 
are not well taken into account within the present approximation to derive a closed RG procedure.

\section{ Analysis of the renormalization rules }

\label{sef_rgrules}

\subsection{ Block Self-dual RG rules }

Let us summarize the output of the previous section : 
once all the odd spins have been eliminated, the Floquet dynamics of the even spins
is described by the evolution operator of Eq. \ref{nr} with the renormalized couplings and fields
satisfying Eqs \ref{jr} and \ref{hr} 
\begin{eqnarray}
\tan(T_0 J^R_{2n-2,2n}) 
&& = \tau_{2n-1}   \frac{\tan(T_0 J_{2n-2} )   \vert  \tan ( T_0 J_{2n-1})  \vert }
{ \sqrt{\tan^2 ( T_0 J_{2n-1})+\tan^2 (T_1 h_{2n-1}) +  \tan^2 (T_1 h_{2n-1}) \tan^2 ( T_0 J_{2n-1})  }}
\nonumber \\
\tan(T_1 h^R_{2n}) && =  
\frac{ \tan(T_1 h_{2n} )   \vert  \tan (T_1 h_{2n-1}) \vert }
{ \sqrt{\tan^2 ( T_0 J_{2n-1})+\tan^2 (T_1 h_{2n-1}) +  \tan^2 (T_1 h_{2n-1}) \tan^2 ( T_0 J_{2n-1})  }}
\label{hrjrblock}
\end{eqnarray}
This defines a closed mapping between the tangents of the phases associated to $(T_0 J_n)$ and $(T_1 h_n)$,
up to the signs $\tau_{2n-1} =\pm 1 $ that label the emergent local integrals of motions associated to the pairs $(2n-1,2n)$.

The corresponding order of the eigenstates can be analyzed via the ratios
\begin{eqnarray}
\rho_n \equiv \frac{ \vert \tan(T_0 J_{n-1}) \vert}{ \vert \tan(T_1 h_{n}) \vert } && 
\label{defratio}
\end{eqnarray}
that satisfy the very simple multiplicative rule
\begin{eqnarray}
\rho_{2n}^R \equiv \frac{ \vert \tan(T_0 J^R_{2n-2,2n}) \vert }{ \vert \tan(T_1 h^R_{2n}) \vert } && 
=    \frac{  \vert \tan(T_0 J_{2n-2} ) \tan ( T_0 J_{2n-1})  \vert } {\vert  \tan (T_1 h_{2n-1})  \tan(T_1 h_{2n} )  \vert  } = \rho_{2n-1} \rho_{2n}
\label{rgratio}
\end{eqnarray}
Equivalently, their logarithms satisfy the additive rule
\begin{eqnarray}
\log \rho_{2n}^R = \log \rho_{2n-1} + \log \rho_{2n}
\label{rgratiolog}
\end{eqnarray}

The location of the critical point between the Paramagnetic Phase (where the renormalized ratios $\rho^R$ flow towards zero) and the Spin-Glass Phase
(where the renormalized ratios $\rho^R$ flow towards infinity)
 is thus given by the following criterion in terms of the disorder average denoted by the overline
\begin{eqnarray}
{\rm Criticality }:  \ \ \  0 = \overline{ \log \rho_n  }  = \overline{ \log \vert \tan(T_0 J_{n}) \vert }   -  \overline{ \log  \vert \tan(T_1 h_{n}) \vert }
\label{critifloquet}
\end{eqnarray}
In addition, the renormalization rule of Eq. \ref{rgratiolog} yields that the critical point corresponds to an Infinite Disorder Fixed Point
with the activated exponent $\psi=1/2$, the typical correlation exponent $\nu_{typ}=1/2$ and the averaged correlation exponent $\nu_{av}=1/2$
exactly as for the Fernandez-Pacheco self-dual procedure for the time independent random quantum Ising chain \cite{nishiRandom,us_pacheco,us_renyi,us_watermelon}.

\subsection {Strong Disorder RG procedure }

\label{sec_strong}

Since the Block Self-dual RG rules discussed above points towards an Infinite Disorder Fixed Point,
the critical properties are expected to be described exactly in the asymptotic regime by
the appropriate Strong Disorder RG rules \cite{fisher,strong_review}.
Here one does not need to do new computations, since one can derive them 
as a special limit from the block self-dual RG rules given above.
The idea is that one wishes to eliminate only one degree of freedom at each step (instead of the $\frac{N}{2}$ odd spins in parallel)
with the following procedure :

(i) one chooses the maximum among the variables $( \vert \tan(T_0 J_{n}) \vert ,   \vert \tan(T_1 h_{n}) \vert )$

(ii) if the maximum corresponds to  $\vert \tan (T_1 h_{n_0} )\vert $, the corresponding spin $\sigma_{n_0}$ is removed and replaced by
the renormalized coupling between its two neighbors
\begin{eqnarray}
\tan(T_0 J^R_{n_0-1,n_0+1}) \simeq \tau_{n_0}   \frac{\tan(T_0 J_{n_0-1} )   \vert  \tan ( T_0 J_{n_0})  \vert }
{ \vert \tan (T_1 h_{n_0}) \vert }
\label{jrstrong}
\end{eqnarray}
that can be obtained from the rules of Eq. \ref{hrjrblock} for the case $n_0=2n-1$ within the approximation for the denominator
$\sqrt{\tan^2 ( T_0 J_{2n-1})+\tan^2 (T_1 h_{2n-1}) +  \tan^2 (T_1 h_{2n-1}) \tan^2 ( T_0 J_{2n-1})  }  \simeq \vert \tan (T_1 h_{2n-1} )\vert$.

(iii) if the maximum corresponds to  $\vert \tan (T_0 J_{n_0} ) \vert$, one replaces the pair $(\sigma_{n_0},\sigma_{n_0+1})$  by a single
renormalized spin with the renormalized transverse field
\begin{eqnarray}
\tan(T_1 h^R_{n_0+1}) && \simeq
\frac{ \tan(T_1 h_{n_0+1} )   \vert  \tan (T_1 h_{n_0}) \vert }
{ \vert \tan ( T_0 J_{n_0}) \vert }
\label{hrstrong}
\end{eqnarray}
that can be obtained from the rules of Eq. \ref{hrjrblock} for the case $n_0=2n-1$ within the approximation for the denominator
$\sqrt{\tan^2 ( T_0 J_{2n-1})+\tan^2 (T_1 h_{2n-1}) +  \tan^2 (T_1 h_{2n-1}) \tan^2 ( T_0 J_{2n-1})  }  \simeq \vert \tan (T_0 J_{2n-1} )\vert$.

\subsection { Limit of small period $T=T_0+T_1 \to 0$ }

In the limit of small period $T \to 0$, the tangents can be linearized and read in terms of the averaged couplings of Eq. \ref{jnav} and \ref{hnav}
of the averaged Hamiltonian of the Magnus expansion of Eq. \ref{havquantumising}
\begin{eqnarray}
\tan(T_0 J_{n} ) && \simeq T_0 J_n = T J_n^{tav}
\nonumber \\
\tan (T_1 h_n) && \simeq T_1 h_n = T h_n^{tav}
\label{tanlinear}
\end{eqnarray}
The RG rules of Eq. \ref{hrjrblock} then become in the limit $T \to 0$
\begin{eqnarray}
 J^{tavR}_{2n-2,2n} 
&& = \tau_{2n-1}   \frac{ J^{tav}_{2n-2}    \vert   J^{tav}_{2n-1}  \vert }
{ \sqrt{ (J^{tav}_{2n-1})^2+ (h^{tav}_{2n-1})^2 }}
\nonumber \\
h^{tavR}_{2n} && =  
\frac{  h^{tav}_{2n}    \vert  h^{tav}_{2n-1} \vert }
{\sqrt{ (J^{tav}_{2n-1})^2+ (h^{tav}_{2n-1})^2 }  }
\label{hrjrhamiltonianlimit}
\end{eqnarray}
These rules coincide with the Fernandez-Pacheco self-dual procedure for the time independent random quantum Ising chain \cite{nishiRandom,us_pacheco,us_renyi,us_watermelon}. 

In this limit of small period $ T \to 0$, the criticality condition of Eq. \ref{critifloquet} yields the usual criterion 
\cite{pfeuty,fisher,strong_review} for the quantum Ising chain of Eq. \ref{havquantumising}
\begin{eqnarray}
{\rm Criticality }:  \ \ \  0 =  \overline{ \log \vert T_0 J_{n} \vert }   -  \overline{ \log  \vert T_1 h_{n} \vert } = 
\overline{ \log \vert  J_{n}^{tav} \vert }   -  \overline{ \log  \vert  h_{n}^{tav} \vert }
\label{critiusual}
\end{eqnarray}

\section{ Conclusion}

\label{sec_conclusion}

In this paper, we have considered a model of periodic quenches between
 two random quantum spin chain Hamiltonians, where the time-evolution operator over one period
 reduces to the product of two simple transfer matrices.
 We have proposed to construct the corresponding localized eigenstates
 via some Block-self-dual renormalization procedure. 
We have also discussed the alternative Strong Disorder Renormalization procedure,
 that generalizes the RSRG-X procedure
 to construct the localized eigenstates of time-independent Hamiltonians.
For the specific model that we have considered, we have obtained that
 the transition between Spin-Glass and Paramagnetic eigenstates 
is described by the Fisher Infinite Disorder Fixed Point \cite{fisher,strong_review},
whose location is determined by Eq. \ref{critifloquet} that replaces the usual criterion of Eq. \ref{critiusual}
concerning the time-independent random quantum Ising chain.

Our main conclusion is that this idea of real-space renormalization to characterize
 the localized eigenstates of the Floquet dynamics in random systems
 provides an interesting alternative point of view with respect
 to the usual Magnus expansion. 
This approach
 can be applied to other models and to higher dimensions $d\rangle1$. 
Indeed for time-independent random Hamiltonians, the real-space renormalization approach
 has been extended to higher-dimensions, both within the Strong Disorder framework
  \cite{fisherreview,motrunich,lin,karevski,lin07,yu,kovacsstrip,kovacs2d,kovacs3d,kovacsentropy,kovacsreview},
 or within the Block self-dual framework \cite{nishiRandom,us_pacheco} : 
the renormalization rules cannot be solved explicitly anymore, 
but they can be implemented numerically on large systems. 
The localized phases of Floquet dynamics for random spin models in $d\rangle1$ could thus
 be studied similarly via the numerical application of the real-space renormalization rules.

\end{document}